\documentclass[floats,floatfix,amssymb,prd,twocolumn,superscriptaddress,nofootinbib]{revtex4}

\usepackage{graphicx,epsf, epsfig, amssymb}
\usepackage{bm}
\usepackage{longtable,tabularx}
\usepackage{xcolor}
\usepackage[breaklinks]{hyperref}
\usepackage{textcomp}
\usepackage{amsfonts,amsmath,amssymb,mathrsfs,gensymb}
\usepackage{natbib}
\usepackage{array}
\usepackage{multirow}
\usepackage{rotating,array}
\usepackage{graphicx}

\newcommand{\A}{{\scriptscriptstyle{A}}}
\newcommand{\B}{{\scriptscriptstyle{B}}}
\newcommand{\C}{{\scriptscriptstyle{C}}}
\newcommand{\D}{{\scriptscriptstyle{D}}}
\newcommand{\E}{{\scriptscriptstyle{E}}}
\newcommand{\F}{{\scriptscriptstyle{F}}}
\newcommand{\G}{{\scriptscriptstyle{G}}}
\newcommand{\I}{{\scriptscriptstyle{I}}}
\newcommand{\J}{{\scriptscriptstyle{J}}}
\newcommand{\M}{{\scriptscriptstyle{M}}}

\newcommand{\new}[1]{\textcolor{black}{#1}}

\newcommand*{\defeq}{\mathrel{\vcenter{\baselineskip0.5ex \lineskiplimit0pt
                     \hbox{\scriptsize.}\hbox{\scriptsize.}}}%
                     =}

\newcounter{count}

\begin{document}

\title{The high-energy collision of black holes in higher dimensions}

\author{Ulrich Sperhake}
\email{U.Sperhake@damtp.cam.ac.uk}
\affiliation{Department of Applied Mathematics and Theoretical Physics, Centre for Mathematical Sciences, University of Cambridge, Wilberforce Road, Cambridge CB3 0WA, United Kingdom}
\affiliation{Department of Physics and Astronomy, The University of
Mississippi, University, Mississippi 38677, USA}
\affiliation{California Institute of Technology, Pasadena, California 91125, USA}

\author{William Cook}
\affiliation{Department of Physics, Princeton University, Jadwin Hall, Washington Road, Princeton, New Jersey 08544, USA}

\author{Diandian Wang}
\email{diandian@physics.ucsb.edu}
\affiliation{Department of Physics, University of California, Santa Barbara, California 93106, USA}

\date{\today}

\begin{abstract}

We compute the gravitational wave energy $E_{\rm rad}$ radiated in head-on
collisions of equal-mass, nonspinning black holes in up to $D=8$
dimensional asymptotically flat spacetimes
for boost velocities $v$ up to about $90\,\%$ of the speed of light.
We identify two main regimes: Weak radiation at velocities
up to about $40\,\%$ of the speed of light, and exponential growth
of $E_{\rm rad}$ with $v$
at larger velocities. Extrapolation to the speed of light predicts
a limit of $12.9\,\%$ $(10.1,~7.7,~5.5,~4.5)\,\%$ 
of the total mass that is lost in
gravitational waves in $D=4$ $(5,\,6,\,7,\,8)$ spacetime dimensions.
In agreement with perturbative calculations,
we observe that the radiation is minimal for small but finite
velocities, rather than for collisions starting from rest.
Our computations support
the identification of regimes with super Planckian curvature
outside the black-hole horizons reported by Okawa, Nakao, and Shibata
[Phys.~Rev.~D {\bf 83} 121501(R) (2011)].

\end{abstract}

\maketitle

\section{Introduction}

The study of general relativity (GR) in more than four spacetime dimensions has
many motivations; in the search for a theory of quantum gravity, often
investigated in the context of string theory; in the study of the gauge/gravity
duality relating higher-dimensional GR to lower-dimensional conformal field
theories; and in the insights provided by the interesting behavior of GR in
the limit that $D$ tends to $\infty$ to name but three.

One particular application of interest of higher-dimensional GR is in the
context of TeV gravity scenarios, proposed to explain the hierarchy problem
between the electroweak scale and Planck scale. In such theories there 
exist large extra dimensions of size $\mathcal{O}(\rm mm)$
into which gravity can leak, tuning down the Planck
scale to $\mathcal{O}(1)$ TeV
\cite{ArkaniHamed:1998rs,Antoniadis:1990ew,Antoniadis:1998ig} . It has been
proposed that in such scenarios, trans-Planckian particle collisions could
result in black-hole (BH) formation in events observed in cosmic rays, or at
high-energy particle colliders such as the LHC
\cite{Banks:1999gd,Giddings:2001bu,Dimopoulos:2001hw}. From a gravitational perspective, it is
proposed that the collision of two highly boosted BHs should approximate such a
collision.

In $D=4$ spacetime
dimensions the problem of colliding BHs, and the study of the radiated
energy has been extensively studied, with the advent of numerical relativity
providing the opportunity to fully study the nonlinear behavior at the moment
of the BH merger.
Prior to numerical approaches, well-known results of Hawking
\cite{Hawking:1971tu} and Penrose \cite{Penrose1974}, detailed in \cite{D'Eath:1992hb,Eardley:2002re}, estimated the upper bound on radiation
from head-on collisions to be 29\% of the total Arnowitt-Deser-Misner
(ADM) mass \cite{Arnowitt:1962hi} of the system,
followed later by the perturbative results of D'Eath and Payne considering the
case of colliding Aichelburg-Sexl shockwaves, providing an estimate of 16.4\%
in the limit that two colliding black holes were boosted to the speed of light
\cite{D'Eath:1976ri,D'Eath:1992hb,D'Eath:1992hd,D'Eath:1992qu}. Similar
calculations have been performed in higher dimensions, which find that the
radiated energy in gravitational waves (GWs) as a function of $D$ should vary
as $\frac{1}{2}-\frac{1}{D}$ \cite{Herdeiro:2011ck,Coelho:2012sy,Coelho:2012sya,Coelho:2014gma}.
See also \cite{Eardley:2002re} for bounds regarding the radiated energy
in BH formation by particle collisions in higher dimensions.
Early numerical
results by Anninos {\em et al.}
in $D=4$ \cite{Anninos:1993zj} considering head-on
collisions from rest have since been followed by an exploration of high-energy
BH collisions; probing the radiated energy for head-on collisions for
equal \cite{Sperhake:2008ga} and unequal mass \cite{Sperhake:2015siy}, with
results independently verified in \cite{Healy:2015mla}, finding that
approximately 13\% of the ADM mass is lost in GW emission.
Further to
this, grazing collisions and collisions of spinning BHs were studied in
\cite{Shibata:2008rq,Sperhake:2009jz,Sperhake:2012me}
with the grazing collisions exhibiting
zoom-whirl behavior \cite{Glampedakis:2002ya,Pretorius:2007jn} and resulting
in near extremal Kerr BHs radiating approximately 50\% of the ADM mass of the
spacetime. The study of the collision of spinning black holes provided evidence
for the so-called matter-does-not-matter conjecture, that in the limit of
high boosts, as kinetic energy dominates, the internal structure of the
colliding objects, such as their spins, ceases to affect the outcome of the
collision, supported also by simulations of boosted collisions of fluid balls
and boson stars \cite{Choptuik:2009ww,East:2012mb,Rezzolla:2012nr}.

To more accurately model the high-energy interactions of TeV gravity scenarios,
it is necessary to explore such boosted BH collisions in more than
four spacetime dimensions.
Since the breakthrough in numerical relativity
\cite{Pretorius:2005gq,Baker:2005vv,Campanelli:2005dd}, it has been possible to
use numerical techniques to explore a variety of questions about fundamental
physics \cite{Cardoso:2014uka,Barack:2018yly}. In particular, the study of
higher-dimensional spacetimes with numerical relativity has been very fruitful, allowing the
investigation of the stability of black objects 
\cite{Shibata:2010wz,Lehner:2010pn,Figueras:2015hkb,Figueras:2017zwa,Bantilan:2019bvf}, as well as simulations of the collisions of black holes from rest
\cite{Witek:2010xi,Cook:2017fec} and with initial momentum
\cite{Okawa:2011fv,Zilhao:2011yc,Cook:2018fxg}. The work of Okawa {\em et al.}
\cite{Okawa:2011fv} in particular has raised the interesting proposal that in
grazing collisions in higher dimensions, super-Planckian curvature can be
formed in regions outside of an event horizon.

In this paper we report on head-on, boosted collisions of nonspinning,
Schwarzschild-Tangherlini BHs in spacetime dimensions $D=4,\,\ldots,\,8$,
and investigate the energy radiated in the emission of gravitational waves. We
also study regions of high curvature that appear to form outside of a common
horizon. In Sec.~\ref{sec:comp} we introduce the computational framework
used to perform the simulations of these collisions. In
Sec.~\ref{sec:codetest} we present the results from tests of our numerical code,
followed by the results of our simulations in Sec.~\ref{sec:results}. We
present our conclusions in Sec.~\ref{sec:conc} and the calculations
that provide our boosted BH initial data in the Appendix \ref{sec:inidata}. We use units where the speed of light and the
Planck constant are $c=\hbar=1$.

%

\section{Computational framework}\label{sec:comp}
The simulations reported below have been performed with the {\sc lean}
code \cite{Sperhake:2006cy,Sperhake:2007gu} which employs the
Baumgarte-Shapiro-Shibata-Nakamura-Oohara-Kojima (BSSNOK)
\cite{Nakamura:1987zz,Shibata:1995we,Baumgarte:1998te}
formulation of the Einstein equations and the {\em moving puncture}
approach for modeling BHs \cite{Baker:2005vv,Campanelli:2005dd}.
{\sc lean} is based on the {\sc cactus} computational toolkit
\cite{Allen:1999,Cactusweb} and uses mesh refinement provided
by {\sc carpet} \cite{Carpetweb,Schnetter:2003rb}. In this work
we focus on higher-dimensional general relativity and consider
asymptotically flat, $D$-dimensional spacetimes with $SO(D-3)$ isometry, i.e.~rotational symmetry in all but three spatial dimensions. This class of spacetimes includes, among other configurations, the head-on collision of
nonspinning BHs, which are the main subject of our study.

For spacetimes with this symmetry,
there are different approaches to dimensionally reduce
the problem to an effectively three-dimensional
computational domain where a few extra field variables
encode all information about the extra dimensions
\cite{Pretorius:2004jg,Shibata:2010wz,Zilhao:2010sr,
Yoshino:2011zz,Yoshino:2011zza,Zilhao:2013gu};
see also the review \cite{Cardoso:2014uka}. Here we use an approach
sometimes referred to as the {\em modified cartoon} method
which represents a generalization of the cartoon technique developed
for the modeling of axisymmetric spacetimes in $3+1$ codes
in Ref.~\cite{Alcubierre:1999ab}. The specific set of equations
and variables we use are those detailed in \cite{Cook:2016soy}.

The physical analysis of our simulations relies on the computation
of the GW energy emitted during the collisions and
the properties of the remnant BH formed therein. We extract the
GW energy using the numerical implementation
of Ref.~\cite{Cook:2016qnt} which is based on the projections
of the Weyl tensor
\cite{Godazgar:2012zq} analogous to the Newman-Penrose scalars commonly
employed in four-dimensional BH simulations. For the diagnostics
of the remnant BHs, we compute the apparent horizon (AH) using
the higher-dimensional AH finder of Ref.~\cite{Cook:2018fxg}
which is based on the techniques developed in
Refs.~\cite{Gundlach:1997us,Alcubierre:1998rq}.

In previous studies of boosted BH binaries
in four or more spacetime dimensions, we have used conformally
flat initial data of Bowen-York \cite{Bowen:1980yu} type which are
analytic solutions of the momentum constraints and where
the Hamiltonian constraint reduces to a differential equation
for the conformal factor that is conveniently solved in the
so-called {\em puncture} approach \cite{Brandt:1997tf,Ansorg:2004ds}.
This approach generalizes in a natural way to higher dimensions
\cite{Yoshino:2006kc,Zilhao:2011yc} but, in either four or
higher dimensions, these data contain
spurious or ``junk'' gravitational radiation that rapidly increases
with the initial boost and leads to large numerical uncertainties
above $v\gtrsim 0.7$; cf.~Fig.~3 in \cite{Sperhake:2008ga}.
More recently, Healy {\em et al.} \cite{Healy:2015mla}
achieved a reduction of the spurious GW content by using
a nonflat conformal metric with appropriate attenuation
functions, reducing the overall error budget in high-energy collisions
in four dimensions.

Here we use a relatively simple construction of initial data following
the approach of \cite{Okawa:2011fv}, which we find to result in
negligible spurious radiation over the entire parameter range explored.
These data consist of the superposition of boosted Tangherlini
\cite{Tangherlini:1963bw} BHs in isotropic coordinates. This ingredient
is the main change in our present study compared to our previous work and
is described in more detail in the Appendix \ref{sec:inidata}.

\section{Results}
In the limit of a single nonboosted BH, our initial data reduce to the
Tangherlini metric in isotropic coordinates (\ref{eq:Tangherliniiso}),
described by one free parameter $\mu$ that determines
the ADM mass $M$ of the spacetime and the Schwarzschild radius $R_S$
of the BH according to \cite{Emparan:2008eg} [see also Eq.~(\ref{eq:Tangherlini})]
\begin{equation}
  M = \frac{(D-2)\Omega_{D-2}}{16\pi\,G}\mu\,,~~~~~~~~~~R_S^{D-3}=\mu\,.
  \label{eq:MRS}
\end{equation}
Here $\Omega_{D-2}$ denotes the area of the $D-2$ unit sphere. The superposition
of $N$ such BHs initially at rest represents the analog of Brill-Lindquist
\cite{Brill:1963yv} initial data whose ADM mass is, in the limit of
large separations, the sum of the
individual BH masses.

Here we focus on head-on collisions of two equal-mass, nonspinning BHs,
$\mathcal{A}$ and $\mathcal{B}$,
characterized by three parameters: the initial position $x = \pm x_0$,
the number $D$ of spacetime dimensions, and the initial velocity
$v\defeq v_{\mathcal{B}}=-v_{\mathcal{A}}$ in the center-of-mass frame.
The boost enters
the total mass of the system in the form of a Lorentz factor
$\gamma=1/\sqrt{1-v^2}$
and we accordingly determine the ADM mass
of a binary spacetime from Eq.~(\ref{eq:MRS}) with the substitution
$\mu=\gamma(\mu_{\mathcal{A}}+\mu_{\mathcal{B}})$. In the remainder of this work, we measure energy
in units of the ADM mass, and
length and time in units
of the Schwarzschild radius associated with the {\em rest} mass
of the BH system, i.e.~$R_S=(\mu_{\mathcal{A}}+\mu_{\mathcal{B}})^{1/(D-3)}$.

For our set of BH binaries, we fix $x_0/R_S=10$, vary the number of dimensions
from $D=5$ to $D=8$ and consider initial boost velocities up to a
$D$-dependent maximal velocity, $v_{\rm max}=0.91~(0.85,~0.8,~0.7)$ in
$D=5~(6,~7,~8)$. The limitations in the velocity range arise from
achieving numerically stable evolutions of the
increasingly steep gradients of the metric fields encountered
at larger $D$.

For our simulations we have used a grid setup (in units of $R_S$)
\begin{equation}
  \left\{(288,\,144,\,96,\,64)\times(5,\,2.5,\,1.25,\,0.625),~h=1/96\right\}
        \nonumber
\end{equation}
using the notation of Sec.~II~E in \cite{Sperhake:2006cy}.
In the following we first discuss code tests
to calibrate numerical uncertainties and validate the suitability of
our initial data. Next, we present and discuss the results obtained from
our set of simulations.

\subsection{Code tests}\label{sec:codetest}
The initial data constructed according to the procedure of
the Appendix
only satisfy the Einstein constraints if assuming one of
the following limits: (i) large initial
separation $x_0\rightarrow \infty$, (ii) vanishing velocity $v\rightarrow 0$,
or (iii) ultrarelativistic velocities $v\rightarrow 1$ (where we recover the
Aichelburg-Sexl metric \cite{Aichelburg:1970dh} and the gravitational field
of an individual ``hole'' is nonvanishing only on a plane orthogonal to the
direction of motion). An additional mitigating factor arises from the
relatively fast falloff of the metric in higher dimensions.
Nevertheless,
it is imperative to verify that constraint violations do not adversely
affect our results beyond the level of accuracy inherent to the
numerical time evolution of the Einstein equations.
This numerical error is estimated below as about $2.5\,\%$.

\begin{figure}
  \includegraphics[width=0.48\textwidth]{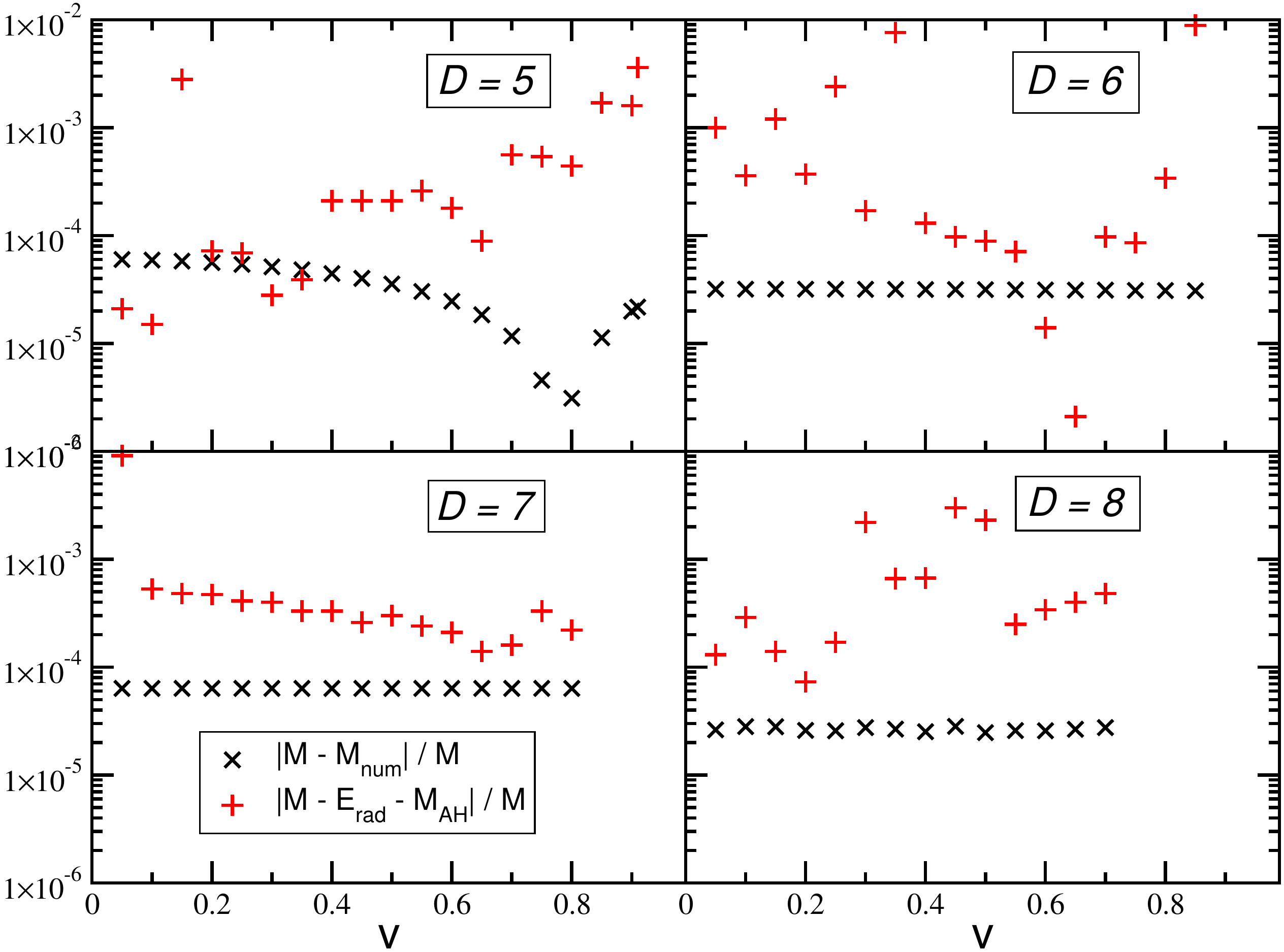}
  \caption{The normalized difference between the analytic and numerical
        ADM mass, $|M-M_{\rm num}|/M$ as obtained from the initial data
        of the Appendix for $D=5$, $6$, $7$ and $8$ and the different
        initial velocities is shown as black $\times$ symbols.
        The red $+$ symbols likewise denote deviations in the
        expected energy balance between the total ADM mass $M$, the horizon mass $M_{\rm AH}$ of the merger remnant BH and the radiated GW energy, i.e. $|M-E_{\rm rad}-M_{\rm AH}|/M$.}
  \label{fig:check}
\end{figure}

We have verified the consistency of our initial data through the following
three tests. First, we compute a numerical estimate $M_{\rm num}$ for the
ADM mass
of the binary initial data
from the metric components [see e.g.~Eq.~(134) in \cite{Cardoso:2014uka}].
This value is compared with the sum
\begin{equation}
  M = \gamma \frac{(D-2)\Omega_{D-2}}{16\pi\,G}(\mu_{\mathcal{A}}
  +\mu_{\mathcal{B}})\,, \nonumber
\end{equation}
which gives the total mass of two BHs with Lorentz factor $\gamma$ in the
large-separation limit. The normalized difference $(M-M_{\rm num})/M$
is displayed as black $\times$ symbols in
Fig.~\ref{fig:check} for our set of simulations.
The excellent agreement (to within $10^{-4}$ or better) demonstrates consistency
of the initial data with the mass energy of a boosted BH binary.

The second test addresses the energy balance throughout the entire
time evolution. Assuming that the spacetime settles down to a stationary
vacuum BH at late times, the ADM mass $M$ has to be equal to the sum of the
postmerger remnant BH mass $M_{\rm AH}$ and the energy
$E_{\rm rad}$ lost in gravitational radiation. The fractional deviation
$(M-E_{\rm rad}-M_{\rm AH})/M$ from energy conservation is shown as
the red $+$ symbols in Fig.~\ref{fig:check} and demonstrates
that energy is conserved in our simulations below the percent level.
The accuracy of this test is limited by the discretization error of the
horizon mass determined in \cite{Cook:2018fxg}
to be about $0.5\,\%$ for the resolution employed here.

For the third consistency test, we have checked the convergence of the
Hamiltonian
constraint [see e.g.~Eq.~(54) in
\cite{Cardoso:2014uka}] for the specific configuration $D=8$, $v=0.6$.
This choice has been motivated by the fact that we generally found it most
difficult to achieve stable and accurate simulations for the case of
\begin{figure}[t]
  \includegraphics[width=0.48\textwidth]{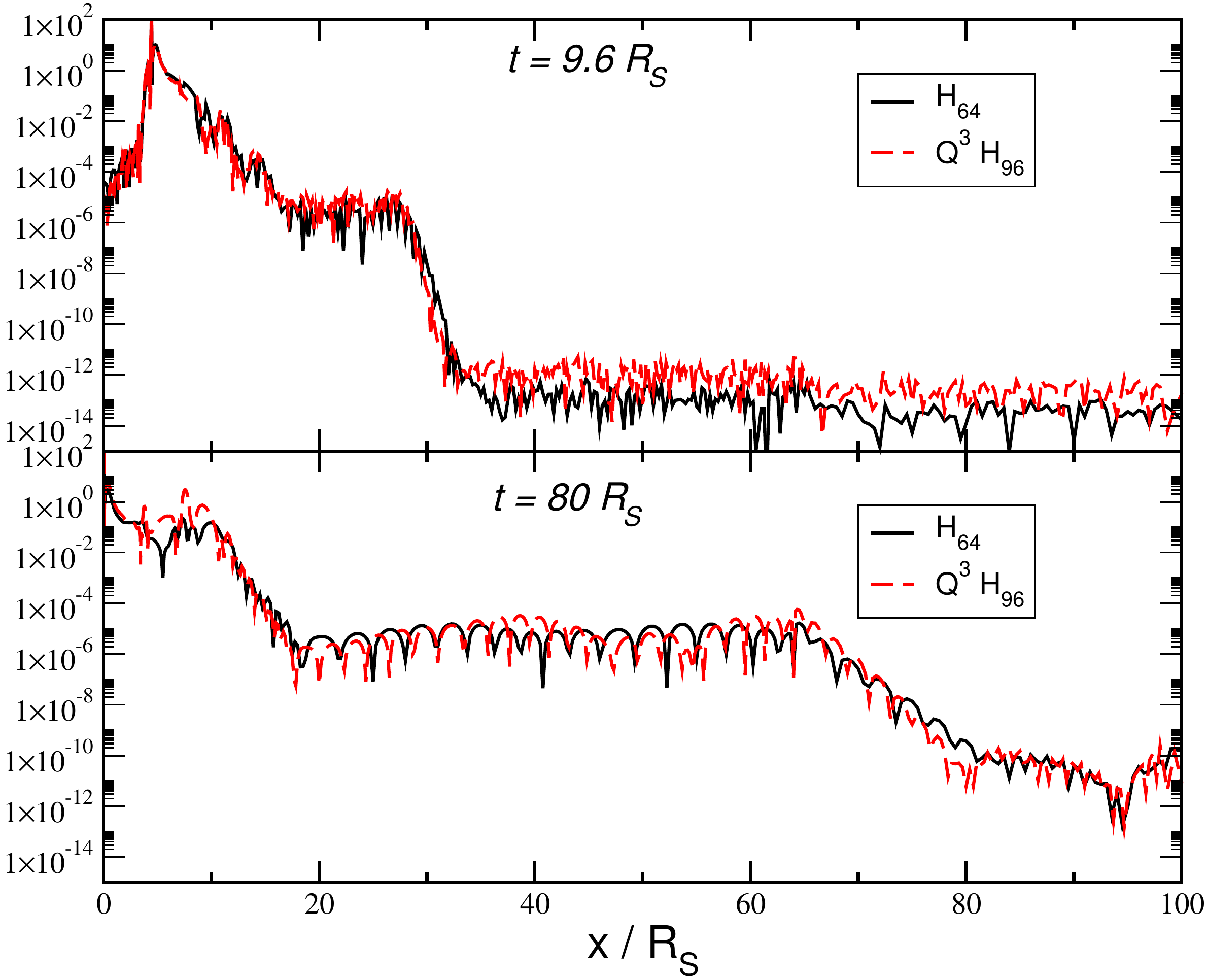}
  \caption{The Hamiltonian constraint
           along the collision axis for a binary with $v=0.6$ in
           $D=8$ dimensions. Note that only the range $x\ge 0$ is
           shown, as the second BH and the range $x<0$ are incorporated
           through reflection symmetry across the origin.
           The times $t=9.6\,R_S$ and $t=80\,R_S$ correspond to
           the infall and postmerger stages of the collision.
           The high-resolution results (red dashed curves) have
           been amplified by a factor $Q^3$, $Q=96/64$, to
           approximately match the low-resolution results,
           indicating convergence at about third order.
           The loss of convergence at $\sim 10^{-13}$ is
           due to roundoff error.}
  \label{fig:constraints}
\end{figure}
moderate to high velocities in $D=8$ dimensions; this is likely due to
the increasingly steep gradients in the metric variables as the
number of dimensions increases. In order to monitor the behavior
of the constraints, we have additionally evolved this configuration
with a grid resolution $h=R_S/64$. Figure \ref{fig:constraints}
displays the violations of the Hamiltonian constraint
along the collision axis at times $t=9.6\,R_S$ (the infall phase
before merger) and $t=80\,R_S$ (in the postmerger ringdown phase).
The high-resolution results have been amplified by a factor $Q^3$
with $Q=96/64$ and the resulting agreement of the curves thus
obtained indicates convergence at about third order, which is in
agreement with the use of fourth- and second-order ingredients in
the discretization \cite{Sperhake:2006cy}. The loss of convergence
at a level of about $10^{-13}$ is due to the roundoff error of the
double precision variables employed in the code. We observe the
same behavior for the momentum constraint, which results in a figure
very similar to Fig.~\ref{fig:constraints}, also showing convergence
at $\approx$ third order.

\begin{figure}
  \includegraphics[width=0.48\textwidth]{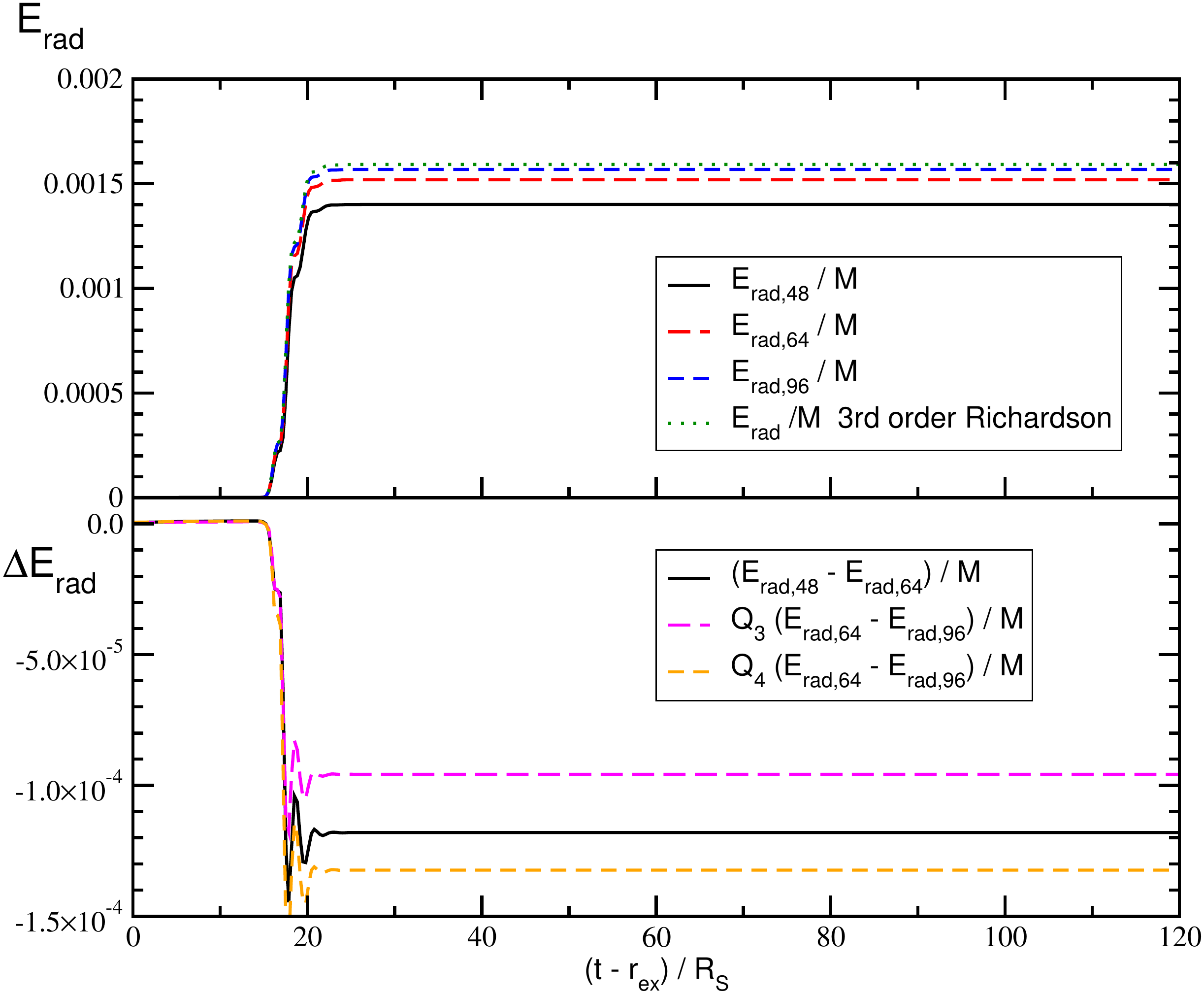}
  \caption{The energy released in gravitational radiation
        in the collision of a BH binary
        starting with $v=0.6$ in $D=8$ dimensions. (Upper panel)
        The results obtained for the different resolutions
        and the prediction obtained from third-order Richardson
        extrapolation. (Lower panel) The differences
        between the individual simulations. The high versus medium
        resolution differences have been amplified by factors
        $Q_3=1.947$ and $Q_4=2.692$ expected for third- and
        fourth-order convergence. The results indicate convergence
        in between and we estimate the uncertainty using the
        more conservative third-order extrapolation to the continuum limit.}
  \label{fig:convErad}
\end{figure}

In order to estimate the discretization error of our results, we have
also studied the convergence of the energy $E_{\rm rad}$ radiated
from this configuration in GWs. We have complemented
the above simulations with a third run at resolution $h=R_S/48$;
unlike the constraints, we do not know the continuum limit of
$E_{\rm rad}$ and, hence, need this extra run. The GW energy $E_{\rm rad}$
is shown in Fig.~\ref{fig:convErad}.
The differences in $E_{\rm rad}$ indicate convergence between third and
fourth order, and we estimate the uncertainty due to discretization using
the more conservative third-order Richardson extrapolation. This yields
a numerical uncertainty of $1.5\,\%$ for the high resolution
($h=R_S/96$). Note that the results of Fig.~\ref{fig:convErad}
contain the spurious gravitational radiation of the initial data,
but this content
is so small that it is not perceptible in the plots,
about $10^{-7}\,M$ for this configuration.
Even though its
contribution can be larger, especially in $D=5$, we have found the
spurious GW content to be orders of magnitude below the discretization
error in all configurations. This is in marked contrast to the major role
of the junk radiation in the error budget of our evolutions of conformally
flat data (see e.g.~\cite{Sperhake:2008ga}) and represents a major
benefit of the superposed BH initial data.

We have analyzed two further sources of numerical uncertainties. First,
the extraction of the gravitational radiation at finite radius incurs
an error which we estimate through extrapolation to infinity using
a series expansion in $1/r$; cf.~Sec.~2 in
\cite{Sperhake:2011zz}. We find this error to be about $1\,\%$ for
$D=5$ and significantly lower for $D>5$. We attribute the small magnitude
of this error once more to the rapid falloff of the metric fields in
higher dimensions, which implies an approximately flat background
metric at smaller radii than in four spacetime dimensions. Finally,
we have varied the initial position of the BHs and find that the
value $x_0=10\,R_S$ is sufficiently large that a further increase
of $x_0$ leads to no significant changes in the results.
In summary, we estimate the relative numerical uncertainty of
our results as about $2.5\,\%$.

\subsection{Numerical results and comparison with analytic calculations}\label{sec:results}
\begin{figure}[t]
  \includegraphics[width=0.48\textwidth]{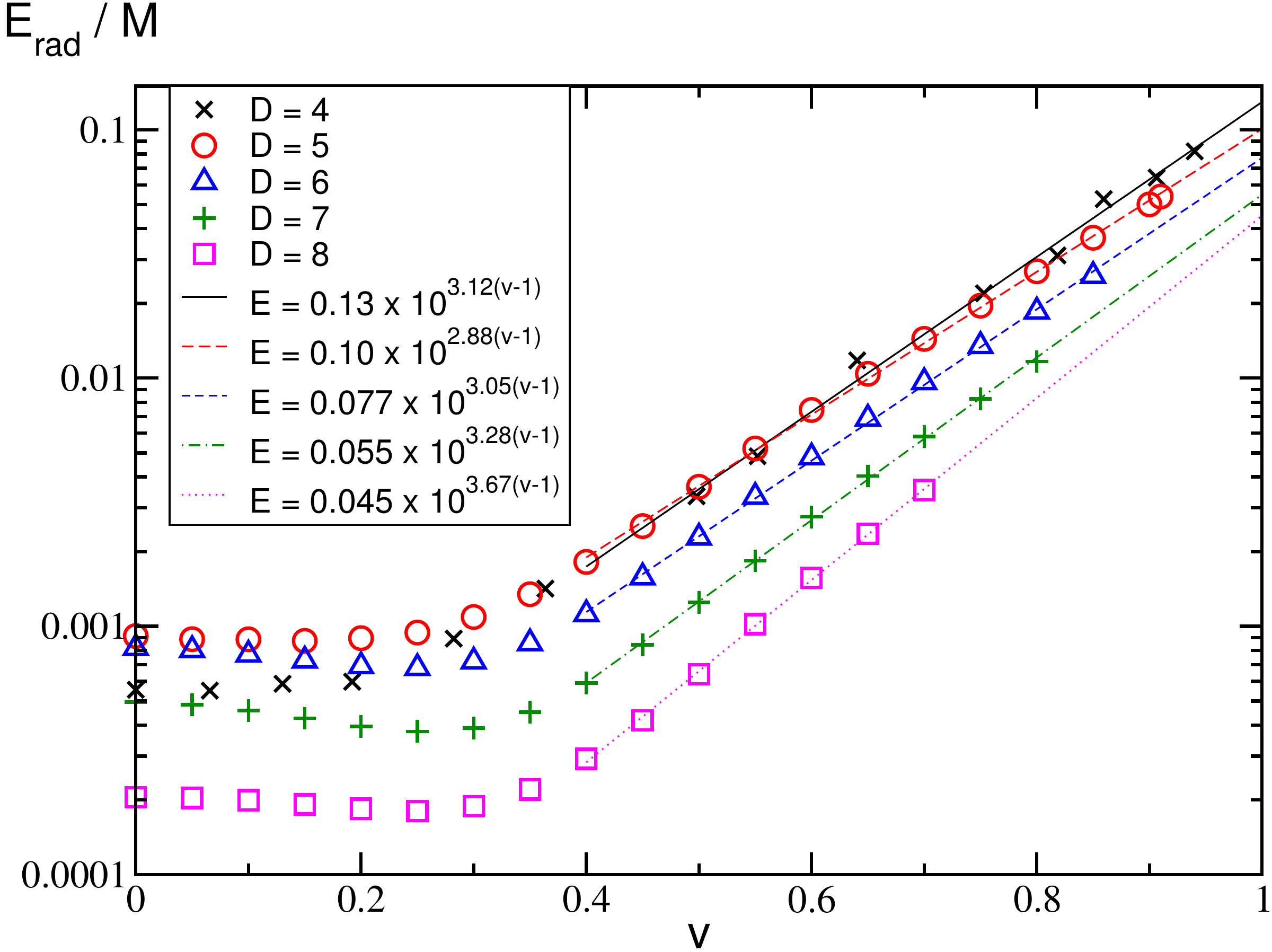}
  \caption{The energy $E_{\rm rad}$
           radiated in gravitational waves from the head-on
           collision of two equal-mass nonspinning BHs with initial
           velocity $v$ in $D$ spacetime dimensions. The fits have
           been computed from data with $v\ge 0.4$ assuming a functional
           relation $\log E_{\rm rad}=a_0+a_1 v$. The results have
           been rewritten to facilitate easy reading of the limit
           $E_{\rm rad}(v\rightarrow 1)$.}
  \label{fig:Eofv}
\end{figure}

The first main result of our work is displayed in Fig.~\ref{fig:Eofv}
which shows the energy radiated in GWs from a binary
with initial boost velocity $v$ in $D$ spacetime dimensions. The
data have been complemented with those obtained in
Ref.~\cite{Sperhake:2008ga} for collisions in $D=4$ dimensions.

For all values $D$, two regimes are distinct in the figure. At velocities
$v\lesssim0.4$, the radiated energy shows mild variation around
the rest-mass limit $E_{\rm rad}(v=0)$ whereas for $v\gtrsim0.4$
the energy grows approximately exponentially with $v$; note the
logarithmic scale on the vertical axis. Contrary to what might
be expected intuitively, the lowest radiation efficiency for a given
$D$ is not always realized in the rest-mass limit. For $D\ge 6$, the function
$E_{\rm rad}(v)$ exhibits a minimum at finite $v\approx 0.25$.
This behavior has in fact already been noticed in point-particle
calculations by Berti {\em et al.} \cite{Berti:2010gx}. In their
Fig.~1, the energy radiated in collisions starting from rest
exceeds that for mild boost velocities for $D\ge 6$; note that,
contrary to our Fig.~\ref{fig:Eofv},
their horizontal axis denotes the number of dimensions while
different symbols mark the velocity. For $D=11$, their rest-mass
case produces even more radiation than the ultrarelativistic
limit. Our dataset does not allow a clear verification of whether
this unexpected phenomenon persists in the comparable
mass limit, but applying fits to our numerical data confirms that
the radiative efficiency in the ultrarelativistic limit
decreases for larger $D$.

For our fits, we have considered only data at $v\ge 0.4$, where
we observe an approximately linear growth of $\log E_{\rm rad}$
with $v$. We therefore apply for each value of $D$ a regression
of the form
\begin{equation}
  \log E_{\rm rad} = a_0 + a_1 v\,.
\end{equation}
It is straightforward to translate the resulting coefficients into
the following notation, where the coefficient in front represents
the limit $E_{\rm rad}(v\rightarrow 1)$,
\begin{eqnarray}
  &E_{\rm rad} = (0.129\pm 0.03) \times 10^{(3.12\pm 0.05)(v-1)}
        ~& \text{ in }D=4 \nonumber \\[5pt]
  &E_{\rm rad} = (0.101\pm0.010) \times 10^{(2.88\pm 0.03)(v-1)}
        ~& \text{ in }D=5 \nonumber \\[5pt]
  &E_{\rm rad} = (0.077\pm0.008) \times 10^{(3.05\pm 0.03)(v-1)}
        ~& \text{ in }D=6 \nonumber \\[5pt]
  &E_{\rm rad} = (0.055\pm0.005) \times 10^{(3.28\pm 0.03)(v-1)}
        ~& \text{ in }D=7 \nonumber \\[5pt]
  &E_{\rm rad} = (0.045\pm0.008) \times 10^{(2.88\pm 0.05)(v-1)}
        ~& \text{ in }D=8\,. \nonumber \\
  && \label{eq:Erad}
\end{eqnarray}
The minor deviation of the result for $D=4$ in this list from the
ultrarelativistic limit reported in \cite{Sperhake:2008ga}
is due to the different functional relations employed in the fits.

\new{It has been noted in Ref.~\cite{Cook:2017fec} that the
overall reduction of the radiated energy with increasing $D$
bears a qualitative resemblance to the decreasing surface
area of the $D$ dimensional unit sphere,
$\mathcal{A}_{D-2}=2\pi^{(D-1)/2}\,/\,\Gamma[(D-1)/2]$.
The $D$ dependence of the radiation efficiency, however, will also
be affected by the
increasingly steep strong-field gradients in larger $D$. These
would be expected to result in more violent interaction, but
also imply that this interaction occurs increasingly close to
merger such that more of the strong-field dynamics are captured
inside the common apparent horizon and cannot radiate to infinity.
The net impact of these competing effects is not obvious,
but our numerical results demonstrate
dominance of those effects {\em reducing} $E_{\rm rad}$.}

We next investigate whether our data confirm the intriguing
observation by Okawa {\em et al.} \cite{Okawa:2011fv} that
\begin{figure*}[t]
  \includegraphics[width=0.48\textwidth]{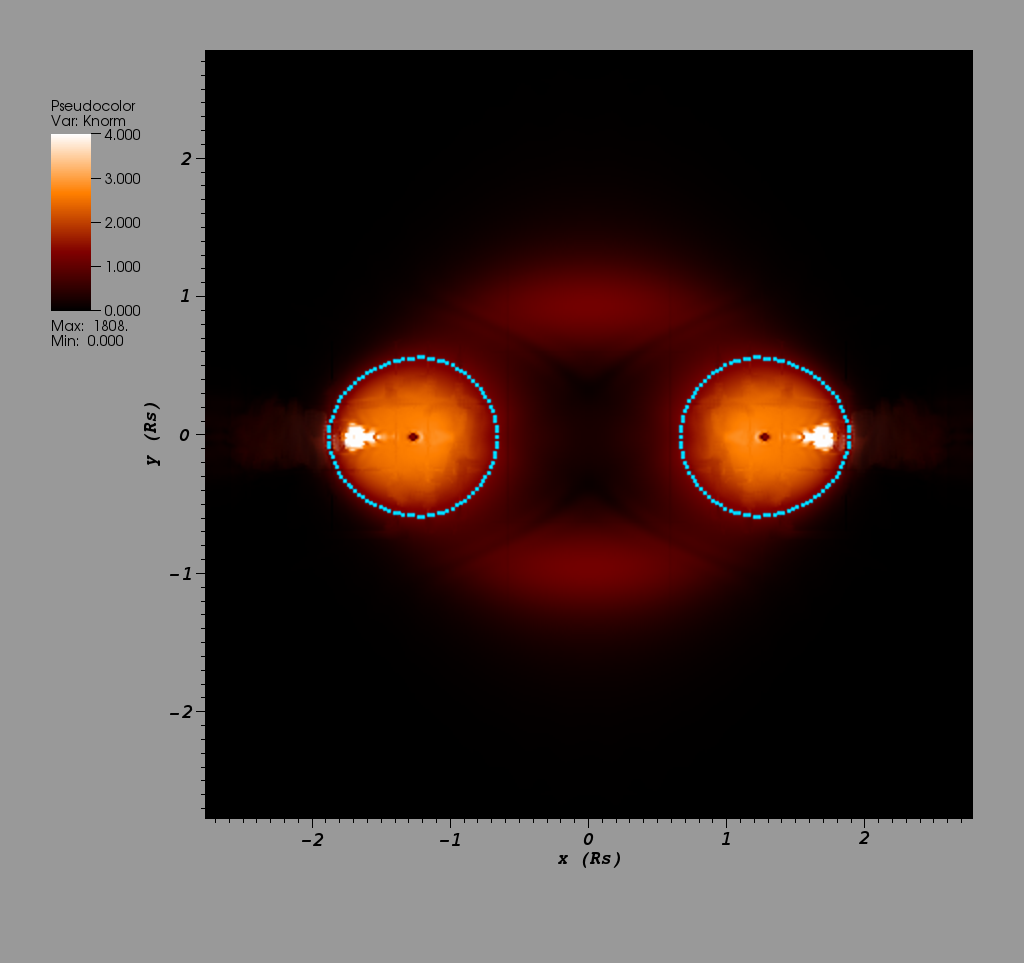}
  \hspace{0.02\textwidth}
  \includegraphics[width=0.48\textwidth]{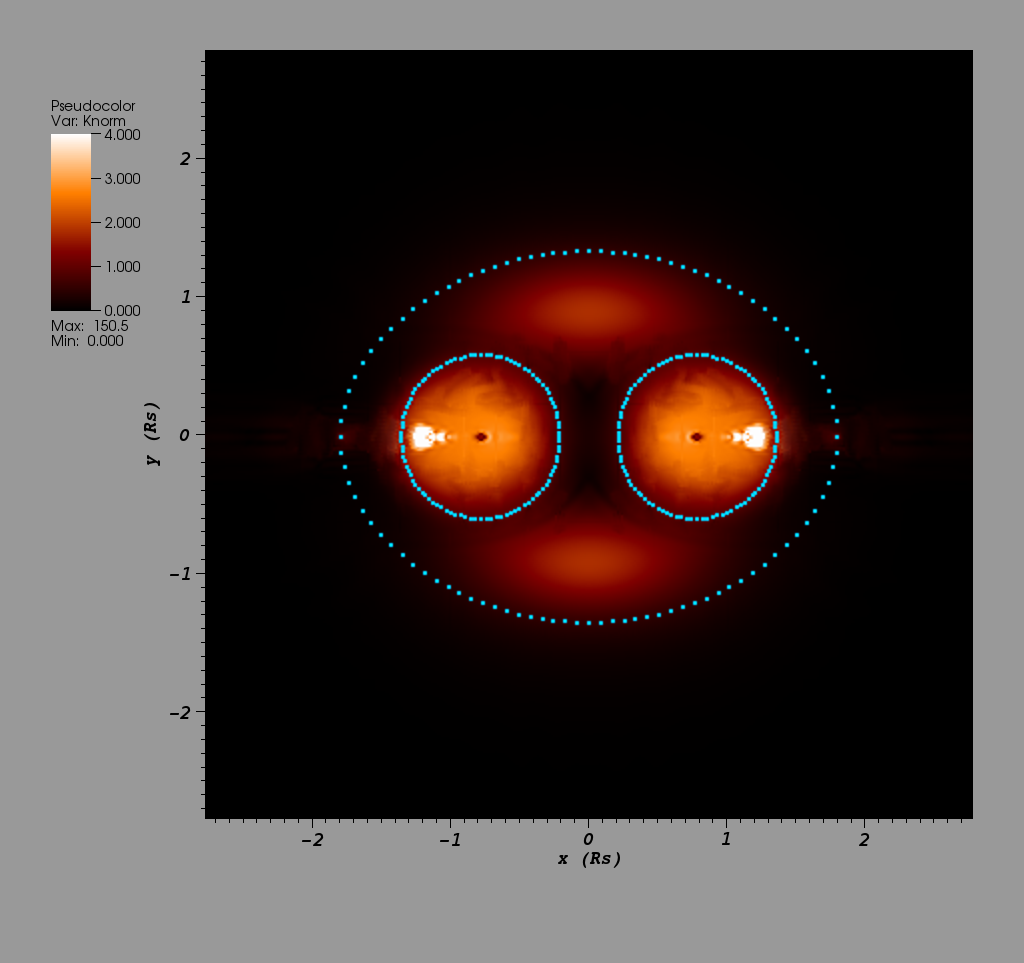}
  \caption{The normalized Kretschmann scalar $\mathcal{K}/\mathcal{K}_p$
           at times $t=12.8\,R_S$ (left) and $t=13.3\,R_S$ (right panel)
           in the collision of a binary with $v=0.85$ in $D=6$
           dimensions. The light-blue lines show the apparent horizon.
           At $t=12.8\,R_S$ two regions where $\mathcal{K}>1$ form, one above and one below the collision axis, indicating
           that super-Planckian curvature may become visible outside
           the BH horizon. At $t=13.3$ a common horizon has formed
           and engulfed this region.}
  \label{fig:KKp}
\end{figure*}
high-energy BH collisions in higher dimensions may form regions
of super-Planckian curvature that are not hidden inside an
event horizon. For this analysis, Okawa {\em et al.}
compute the Kretschmann scalar $\mathcal{K}^2\defeq
R^{\A\B\C\D}R_{\A\B\C\D}$ (where $A,\,B,\,\ldots = 0,\,\ldots,\,D-1$)
and normalize the result with the corresponding value
obtained on the horizon of a BH with a mass equal to the
Planck mass. Their Fig.~2 displays the Kretschmann scalar
thus normalized, and identifies a region of super-Planckian
curvature around the origin and outside the BHs' apparent horizons.

We have explored this phenomenon for our head-on collision
with $v=0.85$ in $D=6$ dimensions. Some care is required in
the comparison, however,
because we use the convention of
\cite{Emparan:2008eg} and write the Einstein equations
as $G_{\A\B}=8\pi G T_{\A \B}$ for all values of $D$, which
mildly differs from the convention of \cite{Okawa:2011fv}.
For our choice, the mass of a Tangherlini BH with mass
parameter $\mu$ is given by Eq.~(\ref{eq:MRS}). We regard
a BH as in the Planckian regime if its Compton wavelength
$1/M_p$ (recall that we set $\hbar=c=1$) is equal to its
horizon radius, i.e.
\begin{eqnarray}
  && \frac{1}{M_p^{D-3}}\overset{!}{=}r_S^{D-3} = \mu
        =\frac{16\pi GM_p}{(D-2)\Omega_{D-2}}\nonumber \\[5pt]
  &\Rightarrow& M_p^{D-2}=\frac{(D-2)\Omega_{D-2}}{16\pi G}\,.
  \label{eq:Mp}
\end{eqnarray}
For $D=6$, we thus obtain for the Planck mass $M_p^4=2\pi/(3G)$.
The Kretschmann scalar on the horizon of a Tangherlini BH
in $D=6$ dimensions is
\begin{equation}
  \mathcal{K}^2 = \frac{240 \mu^2}{r^{10}}\,.
\end{equation}
In this expression we first substitute for $\mu$ in terms
of the BH mass through Eq.~(\ref{eq:MRS}), and then insert for $M$
the Planck mass $M_p$ obtained from Eq.~(\ref{eq:Mp}). The result
gives the Kretschmann scalar on the horizon of
a BH with mass $M_p$ as
\begin{equation}
  \mathcal{K}_p^2 = \frac{180\pi}{G}\,.
\end{equation}
Following Ref.~\cite{Okawa:2011fv} we have computed the normalized
$\mathcal{K}/\mathcal{K}_p$ and show in Fig.~\ref{fig:KKp}
the result in the $xy$ plane;
we recall that this plane is orthogonal to
the $z$ direction, i.e. the quasiradial direction associated with our
rotational isometry \cite{Cook:2016soy}.
The apparent horizon is displayed in the figure
with light blue, dashed curves and contains the regions of highest
curvature. Shortly before we first find a common apparent horizon,
however, two regions of significant curvature $\mathcal{K}
>\mathcal{K}_p$ have formed above and below the collision axis
(left panel in Fig.~\ref{fig:KKp}). This region is eventually
enclosed inside
the common apparent horizon that we first observe at $t=13.3~R_S$
in the right panel.
Our evidence for regions of super-Planckian curvature is less strong
than that presented in \cite{Okawa:2011fv} because our failure
to find an apparent horizon at $t=12.8~R_S$ in the left panel
of Fig.~\ref{fig:KKp} does not prove that an apparent horizon does not
exist. The simulation presented in \cite{Okawa:2011fv}, in contrast,
represents a scattering configuration, which demonstrates more clearly
that a common horizon is not present at the time of super-Planckian
curvature. Nonetheless, our results support their observations, and
indicate that super-Planckian curvature outside a cloaking horizon
may also form in head-on collisions of BHs and in $D>5$.
\new{Theoretically, there is no
reason why super-Planckian curvature outside a BH horizon
cannot occur in $D=4$, but we are not aware of a case where this
has been observed.}

\section{Conclusions}\label{sec:conc}
In this study we have modeled head-on collisions of nonspinning,
equal-mass BH binaries with boost velocities up
to $v_{\rm max}=0.91~(0.85,~0.8,~0.7)$ in $D=5~(6,~7,~8)$ spacetime 
dimensions. By using initial data constructed from superposed
Lorentz boosted Tangherlini BH solutions in isotropic coordinates,
we have managed to significantly reduce the amount of spurious
gravitational radiation as compared with conformally flat initial
data of Bowen-York type. We have verified the suitability of these
initial data by confirming conservation of the total mass energy
and convergence of the Einstein constraints
(Figs.~\ref{fig:check} and \ref{fig:constraints}). We estimate
the relative numerical error of our results to be about $2.5\,\%$
(Fig.~\ref{fig:convErad}, Sec.~\ref{sec:codetest}).
By also including previous results obtained for
boosted head-on collisions in $D=4$ dimensions
\cite{Sperhake:2008ga},
our main findings are summarized as follows.
\begin{list}{\rm{(\alph{count})}}{\usecounter{count}
             \labelwidth1cm \leftmargin1.0cm \labelsep0.4cm \rightmargin0cm
             \parsep0.5ex plus0.2ex minus0.1ex \itemsep0ex plus0.2ex}
  \item Independent of the number of spacetime dimensions,
        we identify two distinct regimes: For initial boosts
        $v\lesssim 0.4$, the radiated GW energy only mildly
        deviates from the limit of collisions starting from rest.
        For $v\gtrsim 0.4$, the radiated energy grows
        approximately exponentially with the velocity parameter
        $v$ (Fig.~\ref{fig:Eofv}).
  \item In agreement with point-particle calculations
        \cite{Berti:2010gx}, we find
        that for $D\ge 6$, the radiated energy as a function
        of initial velocity reaches a local minimum for mild
        but finite boosts $v\approx 0.25$ (Fig.~\ref{fig:Eofv}).
  \item By extrapolating the numerical results to the
        ultrarelativistic limit $v\rightarrow 1$, we
        find that head-on collisions of equal-mass, nonspinning
        BHs radiate $12.9\,\%$, $10.1\,\%$, $7.7\,\%$,
        $5.5\,\%$, $4.5\,\%$ of the total energy in the
        center-of-mass frame, respectively, in $D=4,~5,~6,~7,~8$
        dimensions; cf.~Eq.~(\ref{eq:Erad}).
  \item By computing the Kretschmann curvature scalar for
        head-on collisions in $D=6$ dimensions with initial
        boost $v=0.85$, we identify regions with super-Planckian
        curvature outside the apparent horizon, supporting
        previous numerical results \cite{Okawa:2011fv}
        which show ``visible'' regions of super-Planckian
        curvature in grazing BH collisions in $D=5$.
\end{list}
Our results for the radiated energy demonstrate that high-energy
collisions of BHs can radiate considerable amounts of energy even
in higher dimensions. On the other hand, the values we find
are significantly lower than the remarkable
$E_{\rm rad}/M=\frac{1}{2}-\frac{1}{D}$ formula derived from
first-order perturbative calculations of shock-wave collisions
\cite{Coelho:2012sya,Coelho:2014gma}. In $D=4$, the inclusion
of second-order terms in the perturbative calculations has
lowered the radiation estimate from $E_{\rm rad}^{(1)}=25\,\%$
to $E_{\rm rad}^{(2)}=16.4\,\%$ \cite{D'Eath:1976ri,D'Eath:1992qu}.
First steps have been taken to extend the $D>4$ case to second
order \cite{Coelho:2014gma}. It will be interesting to see if
estimates of the total radiated energy will lead to a similar
reduction and, thus, close the gap between numerical relativity
and shock-wave calculations. Our
numerical results suggest that relatively simple BH production
scenarios based on cross sections derived from the
(higher-dimensional) Schwarzschild radius
\cite{Sirunyan:2018xwt,Giddings:2001bu} would require only
mild modifications by a factor close to unity in order to account
for energy loss through gravitational radiation.

\new{Results in $D=4$ have shown that grazing collisions may emit
gravitational waves more efficiently than the head-on limit; to compute
whether this also holds in higher dimensions is one of the main questions
to be addressed in future work. A further extension of our work
may consider boosted collisions of BHs in
higher-dimensional Lovelock gravity following
the BH solutions and formalism of
Refs.~\cite{Dadhich:2010gu,Dadhich:2012ma,Dadhich:2015lra}.
Such a program, however, might require more investigation to
ensure availability of a well-posed initial-value formulation
\cite{Papallo:2017qvl,Papallo:2019lrl}.}

\section*{ACKNOWLEDGMENTS}
This work was supported by the European Union's H2020 ERC
Consolidator Grant ``Matter and Strong-Field Gravity: New Frontiers
in Einstein’s Theory,'' Grant Agreement No.~MaGRaTh--646597, funding
from the European Union's Horizon 2020 research and innovation
program under Marie Sk\l odowska-Curie Grant
Agreement No.~690904,
COST Action Grant No. CA16104, from STFC Consolidator Grant No.
ST/P000673/1, the SDSC Comet and TACC Stampede2 clusters through
NSF-XSEDE Grant No. PHY-090003, and Cambridge's CSD3 system
through STFC Capital Grant No.~ST/P002307/1 and No.~ST/R002452/1 and STFC
Operations Grant No.~ST/R00689X/1. D.W. acknowledges support from
a Trinity College Summer Research Fellowship. W.C. is supported by
Simons Foundation Grant No.~548512, and the Princeton Gravity
Initiative.

\appendix*
\section{INITIAL DATA FOR BOOSTED BLACK-HOLE BINARIES}
\label{sec:inidata}
In this section we need a wider set of indices to distinguish between
spacetime and spatial, as well as between on- and off-domain spatial indices.
More specifically, we use capital early (middle) latin indices
to cover all spacetime (spatial) dimensions. Lowercase middle
latin indices cover the three spatial directions inside our computational
domain, and early latin indices the extra dimensions outside the computational
domain. Greek indices include time and the on-domain directions.
For $D$ spacetime dimensions, our indices therefore
have the following ranges:
\begin{eqnarray}
  &A,~B,\ldots = 0,\,\ldots,\,D-1\,;~~~
  &I,~J,\ldots = 1,\,\ldots,\,D-1\,; \nonumber \\[5pt]
  &i,~j,\ldots = 1,\,2,\,3\,;~~~
  &a,~b,\ldots = 4,\,\ldots,\,D-1\,. \nonumber \\[5pt]
  &\alpha,~\beta,\ldots = 0,\,1,\,2,\,3\,.
\end{eqnarray}

Our starting point is the Tangherlini metric that describes
a $D$-dimensional, spherically symmetric BH with mass parameter
$\mu$ in radial gauge and polar slicing,
\begin{equation}
  ds^2=-\left(1-\frac{\mu}{R^{D-3}}\right)dt^2
        +\left(1-\frac{\mu}{R^{D-3}}\right)^{-1}dR^2+R^2
        d\omega^2_{D-2}\,,
  \label{eq:Tangherlini}
\end{equation}
where $d\omega^2_{D-2}$ denotes the line element of the $D-2$ sphere.
The metric in isotropic coordinates is obtained by transforming
the radial coordinate according to
\begin{equation}
  R=r\left(1+\frac{\mu}{4r^{D-3}}\right)^{\frac{2}{D-3}}\,,
\end{equation}
which leads to the metric
\begin{eqnarray}
  && ds^2=-\Omega^2 \Psi^{-2} dt^2+\Psi^{\frac{4}{D-3}}(dr^2+r^2
        d\omega_{D-2}^2) \nonumber \\[5pt]
  && \hphantom{ds^2}=-\Omega^2\Psi^{-2}dt^2
        + \Psi^{\frac{4}{D-3}}[ (dx^1)^2 + \ldots
        + (dx^{D-1})^2]\,, \nonumber \\[5pt]
  && \Omega = 1-\frac{\mu}{4r^{D-3}}\,,~~~~~~~~~~
     \Psi = 1+\frac{\mu}{4r^{D-3}}\,.
  \label{eq:Tangherliniiso}
\end{eqnarray}
where $x^1,\,\ldots,\,x^{D-1}$ are standard Cartesian coordinates
with $r^2=(x^1)^2+\ldots +(x^{D-1})^2$.

In the ADM formalism
\cite{Arnowitt:1962hi,York1979}, the spacetime metric
is written in terms of the lapse function $\alpha$,
the shift vector $\beta^I$ and the spatial metric
$\gamma_{IJ}$ according to
\begin{widetext}
\begin{equation}
  g_{AB} = \left( \begin{array}{c|c}
        -\alpha^2+\beta_M\beta^M & \beta_J \\[3pt]
        \hline
        \\[-7pt]
        \beta_I & \gamma_{IJ}
  \end{array}\right)
  =
  \left(
  \begin{array}{c|c|c}
        -\alpha^2+\beta_m \beta^m & \beta_j & 0 \\[5pt]
        \hline
        \\[-7pt]
        \beta_i & \gamma_{ij} & 0 \\[5pt]
        \hline
        \\[-7pt]
        0 & 0 &\gamma_{ww} \delta_{ab}
  \end{array}
  \right)\,,
  \label{eq:gsplit}
\end{equation}
where the first expression is general, and the second accounts
for the simplifications due to
$SO(D-3)$ isometry. For the inverse metric we likewise have
\begin{equation}
  g^{\A\B} = \left(
  \begin{array}{c|c}
    -\alpha^{-2} & \alpha^{-2} \beta^{\J} \\[3pt]
    \hline
    \\[-7pt]
    \alpha^{-2} \beta^{\I} & \gamma^{\I\J} - \alpha^{-2}
        \beta^{\I}\beta{^\J}
  \end{array}
  \right)
  =
  \left(
  \begin{array}{c|c|c}
    \alpha^{-2} & \alpha^{-2} \beta^j & 0 \\[5pt]
    \hline
        \\[-7pt]
    \alpha^{-2}\beta^i & \gamma^{ij}-\alpha^{-2}\beta^i\beta^j & 0 \\[5pt]
    \hline
        \\[-7pt]
    0 & 0 & \gamma^{ww} \delta^{ab}
  \end{array}
  \right)\,.
  \label{eq:invgsplit}
\end{equation}
\end{widetext}
Here $w$ is not an index: $\gamma_{ww}$ and
$\gamma^{ww}=1\,/\,\gamma_{ww}$
merely denote the single extra variable for the metric and inverse metric
needed to describe the geometry in the extra dimensions. We also note
that $\gamma^{ij}$ is the inverse of $\gamma_{ij}$, and $\gamma^{\I\J}$
the inverse of $\gamma_{\I\J}$.

By equating (\ref{eq:gsplit}) and (\ref{eq:invgsplit}) with the
Cartesian metric of Eq.~(\ref{eq:Tangherliniiso}),
we obtain the components for the lapse, shift and spatial metric
\begin{eqnarray}
  &\alpha = \Omega \Psi^{-1}\,,~~~~~~~~& \beta^i = \beta^a = 0\,,
  \nonumber \\[5pt]
  &\gamma_{ij} = \Psi^{\frac{4}{D-3}}\,\delta_{ij}\,,~~~~~~~~&
        \gamma_{ia} = 0\,, \nonumber \\[5pt]
  &\gamma_{ab} = \gamma_{ww}\,\delta_{ab}\,,~~~~~~~~&
        \gamma_{ww} = \Psi^{\frac{4}{D-3}}\,.
  \label{eq:TangherliniisoCart}
\end{eqnarray}
The extrinsic curvature has a more complicated relation to
the metric and also involves derivatives. We use the sign convention
where
\begin{equation}
  K_{\I\J} = -\frac{1}{2\alpha} \left( \partial_0 \gamma_{\I\J}
        -\beta^{\M}\partial_{\M}\gamma_{\I\J}
        -\gamma_{\M\J}\partial_{\I}\beta^{\M}
        -\gamma_{\I\M}\partial_{\J}\beta^{\M}\right)\,.
  \label{eq:KIJ}
\end{equation}
Applied to the Tangherlini metric (\ref{eq:Tangherliniiso}), however,
one directly finds that $K_{\I\J}=0$, because the metric is time independent and has zero shift vector.

The next step in our initial data construction consists of
applying a Lorentz boost to the Tangherlini metric in
Cartesian coordinates.
For this purpose we consider an observer $\mathcal{O}$ in the rest frame
of the BH, and a second observer
$\tilde{\mathcal{O}}$ who
moves with velocity $-v^{\I}$ relative to $\mathcal{O}$.
The transformation between the two frames is given by
\begin{equation}
  x^{\tilde{\A}} = \Lambda^{\tilde{\A}}{}_{\E}x^{\E}
        + x_0^{\tilde{\A}}
        ~~~~~\Leftrightarrow~~~~~
  x^{\E} = \Lambda^{\E}{}_{\tilde{\A}}
        (x^{\tilde{\A}} - x_0^{\tilde{\A}})\,,
\end{equation}
where
\begin{equation}
  \Lambda^{\tilde{\A}}{}_{\E} =
  \left(
  \begin{array}{c|c}
    \Lambda^{\tilde{\alpha}}{}_{\epsilon} & 0 \\[3pt]
    \hline
    \\[-7pt]
    0 & \delta^{\tilde{a}}{}_{e}
  \end{array}
  \right)
  =
  \left(
  \begin{array}{c|c|c}
        \gamma & \gamma v_j & 0 \\[3pt]
        \hline
        \\[-7pt]
        \gamma v^i & \delta^i{}_j+(\gamma-1)\frac{v^iv_j}{|\vec{v}|^2} & 0
        \\[5pt]
        \hline
        \\[-7pt]
        0 & 0 & \delta^{\tilde{a}}{}_{e}
  \end{array}
  \right)\,,
\end{equation}
and its inverse $\Lambda^{\E}{}_{\tilde{\A}}$ is obtained from
the same expression by simply inverting the sign of the velocity $v^i$.
Note that boosts in the extra dimensions are excluded here in order to
preserve the $SO(D-3)$ isometry.
Without loss of generality, we will from now on
set the constant offset $x_0^{\tilde{A}}$
to zero, which merely
implies synchronization of the two observers' clocks when they meet.

The metric components and their derivatives in the two frames
$\mathcal{O}$ and $\tilde{\mathcal{O}}$ are related by
\begin{eqnarray}
  g_{\tilde{\A}\tilde{\B}} &=& \Lambda^{\E}{}_{\tilde{\A}}
        \Lambda^{\F}{}_{\tilde{\B}}\,g_{\E\F}\,, \\[5pt]
  \partial_{\tilde{\C}} g_{\tilde{\A}\tilde{\B}} &=&
        \Lambda^{\G}{}_{\tilde{\C}} \Lambda^{\E}{}_{\tilde{\A}}
        \Lambda^{\F}{}_{\tilde{\B}} \, \partial_{\G}
        g_{\E\F}\,.
\end{eqnarray}
For the eventual calculation, it is convenient to consider separately
in these relations the spacetime components inside our computational
domain and those corresponding to the off-domain directions
$x^a$. This leads to the following transformation rules for the
metric, its inverse and its partial derivatives,
\begin{eqnarray}
  & g_{\tilde{\alpha}\tilde{\beta}} = \Lambda^{\mu}{}_{\tilde{\alpha}}
        \Lambda^{\nu}{}_{\tilde{\beta}} g_{\mu\nu}\,,~~
  & g_{\tilde{a}\tilde{b}} = \delta_{\tilde{a}\tilde{b}}\,g_{ww}\,,
  \nonumber \\[5pt]
  & g^{\tilde{\alpha}\tilde{\beta}} = \Lambda^{\tilde{\alpha}}{}_{\mu}
        \Lambda^{\tilde{\beta}}{}_{\nu}g^{\mu\nu}\,,~~
  & g^{\tilde{a}\tilde{b}} = \delta^{\tilde{a}\tilde{b}}g^{ww}\,,
  \\[5pt]
  &\partial_{\tilde{\gamma}}g_{\tilde{\alpha}\tilde{\beta}}
        = \Lambda^{\lambda}{}_{\tilde{\gamma}}
        \Lambda^{\mu}{}_{\tilde{\alpha}}
        \Lambda^{\nu}{}_{\tilde{\beta}}
        \partial_{\lambda}g_{\mu\nu}\,,~~
  & \partial_{\tilde{\gamma}}g_{\tilde{a}\tilde{b}} =
        \Lambda^{\lambda}{}_{\tilde{\gamma}}
        \delta_{\tilde{a}\tilde{b}}\partial_{\lambda}g_{ww}\,,
        \nonumber
\end{eqnarray}
with all other components and derivatives being manifestly zero.
The ADM variables in the boosted frame $\tilde{\mathcal{O}}$
can then be read off from these expressions through the relations
(\ref{eq:gsplit}), (\ref{eq:invgsplit}) and (\ref{eq:KIJ}),
which hold in exactly the same form in the new coordinates
$x^{\tilde{\alpha}}$. This gives us
\begin{eqnarray}
  && \tilde{\alpha} =
  \big(-g^{\tilde{0}\tilde{0}}\big)^{-1/2}\,,
  ~~~~~\beta_{\tilde{i}}= g_{\tilde{0}\tilde{i}}\,,~~~~~
  \gamma_{\tilde{i}\tilde{j}}=g_{\tilde{i}\tilde{j}}\,,
  \nonumber\\[5pt]
  && \gamma_{\tilde{a}\tilde{b}}= \gamma_{\tilde{w}\tilde{w}}\,
        \delta_{\tilde{a}\tilde{b}}\,,~~~~~
  \gamma_{\tilde{w}\tilde{w}} = g_{\tilde{w}\tilde{w}}
        =g_{ww}\,,
  \nonumber\\[5pt]
  &&K_{\tilde{i}\tilde{j}} = \frac{-1}{2\tilde{\alpha}}
        \left( \partial_{\tilde{0}} \gamma_{\tilde{i}\tilde{j}}
        -\beta^{\tilde{m}}\partial_{\tilde{m}}
        \gamma_{\tilde{i}\tilde{j}}
        -\gamma_{\tilde{m}\tilde{j}}\partial_{\tilde{i}}
        \beta^{\tilde{m}}
        -\gamma_{\tilde{i}\tilde{m}}\partial_{\tilde{j}}\beta^{\tilde{m}}
        \right)\,,
  \nonumber\\[5pt]
  &&K_{\tilde{a}\tilde{b}} = K_{\tilde{w}\tilde{w}}\,
        \delta_{\tilde{a}\tilde{b}}\,.
  \nonumber\\[5pt]
  && K_{\tilde{w}\tilde{w}} = \frac{-1}{2\tilde{\alpha}}
        \left( \partial_{\tilde{0}} \gamma_{\tilde{w}\tilde{w}}
        - \beta^{\tilde{m}}\partial_{\tilde{m}}
        \gamma_{\tilde{w}\tilde{w}}
        -2\gamma_{\tilde{w}\tilde{w}}\frac{\beta^{\tilde{z}}}{\tilde{z}}
        \right)\,.
\end{eqnarray}
Note that we have put a tilde on the index free lapse function
$\tilde{\alpha}$ to distinguish it from the lapse
$\alpha$ in the rest frame $\mathcal{O}$, and
that we have used in the last line the relation \cite{Cook:2016soy}
\begin{equation}
  \lim_{z\rightarrow 0}
        \partial_a \beta^c = \frac{\beta^z}{z} \delta_a{}^c\,.
\end{equation}

This transformation allows us to compute the initial data for
a single boosted BH. For binary data, we compute such a
solution for two BHs $\mathcal{A}$ and $\mathcal{B}$ with
opposite boost velocities $v_{\mathcal{B}}^i=-v_{\mathcal{A}}^i$
and initially located at positions $x_{\mathcal{A}}^i=
-x_{\mathcal{B}}^i$, which gives us the center-of-mass frame
for equal-mass BHs. Following \cite{Sperhake:2005uf}, we construct
superposed binary data from the two individual solutions
according to
\begin{eqnarray}
  \gamma_{\tilde{i}\tilde{j}} &=& \gamma_{\tilde{i}\tilde{j}}^{\mathcal{A}}
        + \gamma_{\tilde{i}\tilde{j}}^{\mathcal{B}}
        - \delta_{\tilde{i}\tilde{j}}\,, \nonumber \\[5pt]
  \hat{K}^{\tilde{i}}{}_{\tilde{j}}
        &=& K^{\tilde{i}}{}_{\tilde{j}}{}^{\mathcal{A}}
        +K^{\tilde{i}}{}_{\tilde{j}} {}^{\mathcal{B}} \,, \nonumber \\[5pt]
  K_{\tilde{i}\tilde{j}} &=& \frac{1}{2} \left(
        \gamma_{\tilde{i}\tilde{m}}
        \hat{K}^{\tilde{m}}{}_{\tilde{j}}
        + \gamma_{\tilde{j}\tilde{m}}
        \hat{K}^{\tilde{m}}{}_{\tilde{i}}\right)\,.
\end{eqnarray}
Instead of superposing the lapse and shift vector in an analogous
way, we initialize the lapse in terms of the conformal factor
of the BSSNOK formulation,
$\tilde{\alpha} = \sqrt{\tilde{\chi}}\,,~~
\tilde{\chi}=(\det \gamma_{\tilde{i}\tilde{j}}
)^{-1/(D-1)}$, and set the initial shift to zero, $\beta^{\tilde{i}}=0$.


\end{document}